\documentclass[10pt]{article}
\usepackage[dvips]{graphicx}

\usepackage{epsfig}
\usepackage{graphicx}
\usepackage{indentfirst}
\usepackage{amsmath}
\usepackage{amsfonts}
\usepackage{amssymb}

\begin{document}
%

\begin{center}
\begin{flushright}\begin{small}    
\end{small} \end{flushright} \vspace{1.5cm}
\huge{Charged Black Holes in Generalized Teleparallel Gravity} 
\end{center}

\begin{center}
{\small  M. E. Rodrigues$^{(a,b)}$\footnote{e-mail: esialg@gmail.com}}, {\small M. J. S. Houndjo$^{(c,d)}$\footnote{e-mail:sthoundjo@yahoo.fr}}, {\small J. Tossa $^{(c)}$\footnote{e-mail: joel.tossa@imsp-uac.org}}, {\small D. Momeni$^{(e)}$\footnote{e-mail: d.momeni@yahoo.com}} and {\small R. Myrzakulov$^{(e)}$\footnote{e-mail:rmyrzakulov@gmail.com}}
\vskip 4mm

(a) \ Faculdade de F\'{\i}sica, Universidade Federal do Par\'{a}, 66075-110, Bel\'em, Par\'{a}, Brazil \\
(b) \ Faculdade de Ci\^{e}ncias Exatas e Tecnologia, Universidade Federal do Par\'{a} - Campus Universit\'{a}rio de Abaetetuba, CEP 68440-000, Abaetetuba, Par\'{a}, Brazil
\\
(c) \ Institut de Math\'{e}matiques et de Sciences Physiques (IMSP) - 01 BP 613 Porto-Novo, B\'{e}nin\\
(d) \ Facult\'e des Sciences et Techniques de Natitingou - Universit\'e de Parakou - Natitingou - B\'{e}nin\\

(e) \  Eurasian International Center for Theoretical Physics - Eurasian National University, Astana 010008, Kazakhstan\\
\vskip 2mm
\end{center}

\begin{center}
                                       Abstract
\end{center}
\hspace{0,5cm} In this paper we investigate charged static black holes in $4D$ for generalized teleparallel models of gravity, based on torsion as the geometric object for describing gravity according to the equivalence principle. As a motivated idea, we introduce a set of non-diagonal tetrads and derive the full system of non linear differential equations. We prove that the common Schwarzschild gauge is applicable only when we study linear $f(T)$ case. We reobtain the Reissner-Nordstrom-de Sitter (or RN-AdS) solution for the linear case of  $f(T)$ and perform a parametric cosmological reconstruction for two nonlinear models. We also study in detail a type of the no-go theorem  in the framework of this modified teleparallel gravity.
\vspace{0,5cm}

Pacs numbers:  04.50. Kd, 04.70.Bw, 04.20. Jb 


\section{Introduction}
Gravity as the old fundamental force in world is described by a gauge theory according to the equivalence principle.  According to Mach's principle, gravity is view as  a deformation of the geometry from the flat spacetime to the curved one. The basic principle reads:
$$
\text{Gravity}\cong\text{Geometry}
$$
\par
The right hand side of the above equation denotes the geometry and it is not clear which kind of geometry (Riemannian or non Riemannian) has to be used. Also, the geometry quantities are scalars and can be constructed from different tensor objects. For example, by using Riemannian geometry, the scalar can be written in the form of any of the following expressions  $g^{ij}R_{ij},R^{ijkl}R_{ijkl},C^{ijkl}C_{ijkl}$ and more. The most simple modification is the replacement of scalar curvature ``R'' by a generic function ``f(R)'', originally proposed by \cite{Buchdahl} and extended in literatures for cosmology and gravity, and also as weak limit of quantum gravity \cite{f(R)}.\par
In recent years, more attention is attached to gravity as effect of the torsion of spacetime, originally introduced in parallel to curvature description \cite{Einstein}, and developed in several years as teleparallel equivalence of gravity (TEGR) \cite{maluf1}. Special forms as  torsion models have been proposed and applied diversely \cite{fT}. 
\par 
Black hole solutions are more compact objects which store informations about entropy on their horizon (originally discovered by Hawking \cite{Hawking}). The horizon of a black hole has a definite topology and temperature, and consequently thermodynamics. In modified TEGR, $f(T)$ has some strange and also interesting features. The main important is what happens at the cosmological level, due to the field equations, and also the fact that the black hole depends on the frame under consideration \cite{Boehmer}. This means that, if we change the tetrad frame from a basic diagonal to a non-diagonal one, performing some kinds of Lorentz transformations, the results are different.  This feature of $f(T)$  theory, and consequently of its related field equations, leads to the non-invariance under Lorentz transformations \cite{manusug1}, absence of evaporation for Nariai black hole in diagonal frame \cite{evaporation} and also existence (non existence) of relativistic stars \cite{manusug2,HamaniDaouda:2011iy}.  Recent observations from solar system orbital motions in order to constrain $f(T)$ gravity have been made and interesting results have been found \cite{Iorio:2012cm}.
\par
Besides neutral black holes there are also charged Maxwell field minimally coupled to gravity. Some early works have studied charged black holes in $f(T)$ theory \cite{Wang:2011xf,saridakis1,saridakis2}. Previously, in \cite{saridakis1}, three dimensional solutions have been investigated in detail and generalized later in D-dimensional frame \cite{saridakis2}. 
\par 
Our main goal in this paper is to study black hole solutions with Maxwell fields in a general non diagonal frame in $4D$. We fundamentally undertake this problem from the point of view of analytic solutions. We choose  a non-diagonal frame and study TEGR, recovering the Reissner-Nordstrom solution. The main and interesting feature here is a theorem according to what, in $f(T)$, we can use the gauge of Schwarzschild where $g_{tt}g_{rr}=-1$. Recently, this gauge has been used as a solver key of obtaining black holes in $f(T)$ \cite{attazadeh}. We will show that $g_{tt}g_{rr}=-1$ is a valid result only for whole manifold $T=T_0$ or when the geometry preserves only TEGR. However, for a generic form of $f(T)$ and with $T\neq T_0$ or $f(T)\neq T+C$, there is no reason to use $g_{tt}g_{rr}=-1$. We classify the equations as a system of coupled non linear differential equations. Moreover, we study the no-go theorem in this theory.
\par
The paper is organized as follows. In sec. \ref{sec2}, we present general derivation of field equations of $f(T)$ gravity and also some discussions about the frames. In sec. \ref{sec3}, we formulate charged black holes in a non-diagonal frame. Sec. \ref{sec4} is devoted to the cosmological reconstruction of two $f(T)$  models. In sec. \ref{sec5} we prove and present the no-go theorem for the generalised $f(T)$ model. We conclude and summarize in sec. \ref{sec6}.
       
\section{\large The field equations from $f(T)$ theory}\label{sec2}

\hspace{0,2cm} In this section we will show how to obtain the equations of motion for $f(T)$ theory and also put out the choice of the matter as an energy-momentum tensor of a spin-1 Maxwell field. 
\par
We start defining the line element as 
\begin{eqnarray}
dS^2=g_{\mu\nu}dx^{\mu}dx^{\nu}=\eta_{ab}\theta^{a}\theta^{b}\label{ele}\;,\\
\theta^{a}=e^{a}_{\;\;\mu}dx^{\mu}\;,\;dx^{\mu}=e_{a}^{\;\;\mu}\theta^{a}\label{the}\;,
\end{eqnarray}
where $g_{\mu\nu}$ is the metric of the space-time, $\eta_{ab}$ the Minkowski metric, $\theta^{a}$ the tetrads and $e^{a}_{\;\;\mu}$ and their inverses $e_{a}^{\;\;\mu}$ the tetrads matrices satisfying the relations $e^{a}_{\;\;\mu}e_{a}^{\;\;\nu}=\delta^{\nu}_{\mu}$ and  $e^{a}_{\;\;\mu}e_{b}^{\;\;\mu}=\delta^{a}_{b}$. The root of the determinant of the metric is given by $\sqrt{-g}=det[e^{a}_{\;\;\mu}]=e$. The Weitzenbok connection is defined by 
\begin{eqnarray}
\Gamma^{\alpha}_{\mu\nu}=e_{i}^{\;\;\alpha}\partial_{\nu}e^{i}_{\;\;\mu}=-e^{i}_{\;\;\mu}\partial_{\nu}e_{i}^{\;\;\alpha}\label{co}\; .
\end{eqnarray}
\par
Through the connection, we can define the components of the torsion and  contorsion tensors  as
\begin{eqnarray}
T^{\alpha}_{\;\;\mu\nu}&=&\Gamma^{\alpha}_{\nu\mu}-\Gamma^{\alpha}_{\mu\nu}=e_{i}^{\;\;\alpha}\left(\partial_{\mu} e^{i}_{\;\;\nu}-\partial_{\nu} e^{i}_{\;\;\mu}\right)\label{tor}\;,\\
K^{\mu\nu}_{\;\;\;\;\alpha}&=&-\frac{1}{2}\left(T^{\mu\nu}_{\;\;\;\;\alpha}-T^{\nu\mu}_{\;\;\;\;\alpha}-T_{\alpha}^{\;\;\mu\nu}\right)\label{cont}\; .
\end{eqnarray}
\par
For facilitating the description of the Lagrangian and the equations of motion, we can define another tensor from the components of the torsion and contorsion tensors, as
\begin{eqnarray}
S_{\alpha}^{\;\;\mu\nu}=\frac{1}{2}\left( K_{\;\;\;\;\alpha}^{\mu\nu}+\delta^{\mu}_{\alpha}T^{\beta\nu}_{\;\;\;\;\beta}-\delta^{\nu}_{\alpha}T^{\beta\mu}_{\;\;\;\;\beta}\right)\label{s}\;.
\end{eqnarray}
Now, defining the torsion scalar 
\begin{equation}
T=T^{\alpha}_{\;\;\mu\nu}S_{\alpha}^{\;\;\mu\nu}\label{t1}\,\,\,,
\end{equation}
one can write the Lagrangian of the $f(T)$ theory, coupled with the matter, as follows 
\begin{equation}
\mathcal{L}=ef(T)+\mathcal{L}_{Matter}\;.\label{lagran}
\end{equation}
The principle of least action leads to the Euler-Lagrange equations. In order to use these equations, we first write the quantities
\begin{eqnarray}
&&\frac{\partial\mathcal{L}}{\partial e^{a}_{\;\;\mu}}=f(T)ee_{a}^{\;\;\mu}+ef_{T}(T)4e_{a}^{\;\;\alpha}T^{\sigma}_{\;\;\nu\alpha}S_{\sigma}^{\;\;\mu\nu}+\frac{\partial\mathcal{L}_{Matter}}{\partial e^{a}_{\;\;\mu}}\;,\label{1}\\
&&\partial_{\alpha}\left[\frac{\partial \mathcal{L}}{\partial (\partial_{\alpha}e^{a}_{\;\;\mu})}\right]=-4f_{T}(T)\partial_{\alpha}\left(ee_{a}^{\;\;\sigma}S_{\sigma}^{\;\;\mu\nu}\right)-4ee_{a}^{\;\;\sigma}S_{\sigma}^{\;\;\mu\alpha}\partial_{\alpha}T\,f_{TT}(T)\nonumber\\
&&+\partial_{\alpha}\left[\frac{\partial \mathcal{L}_{Matter}}{\partial (\partial_{\alpha}e^{a}_{\;\;\mu})}\right]\label{2}\;,
\end{eqnarray}
where $f_{T}(T)=df(T)/dT$ and $f_{TT}(T)=d^2f(T)/dT^2$ denote the first and second derivatives of the algebraic function $f(T)$ with respect to the torsion scalar $T$, respectively. The equations of Euler-Lagrange are given by
\begin{eqnarray}\label{EL}
\frac{\partial\mathcal{L}}{\partial e^{a}_{\;\;\mu}}-\partial_{\alpha}\left[\frac{\partial \mathcal{L}}{\partial (\partial_{\alpha}e^{a}_{\;\;\mu})}\right]=0\;.\label{joas}
\end{eqnarray}
By multiplying (\ref{joas}) by $e^{-1}e^{a}_{\;\;\beta}/4$, one gets 
\begin{eqnarray}
S_{\beta}^{\;\;\mu\alpha}\partial_{\alpha}T\,f_{TT}(T)+\left[e^{-1}e^{a}_{\;\;\beta}\partial_{\alpha}\left(ee_{a}^{\;\;\sigma}S_{\sigma}^{\;\;\mu\alpha}\right)+T^{\sigma}_{\;\;\nu\beta}S_{\sigma}^{\;\;\mu\nu}\right]f_{T}(T)+\frac{1}{4}\delta^{\mu}_{\beta}f(T)=4\pi \mathcal{T}^{\mu}_{\beta}\label{em}\;,
\end{eqnarray}
where the energy momentum tensor is given by 
\begin{eqnarray}
\mathcal{T}^{\mu}_{\beta}=-\frac{e^{-1}e^{a}_{\;\;\beta}}{16\pi}\left\{ \frac{\partial \mathcal{L}_{Matter}}{\partial e^{a}_{\;\;\mu}}-\partial_{\alpha}\left[\frac{\partial \mathcal{L}_{Matter}}{\partial (\partial_{\alpha}e^{a}_{\;\;\mu})}\right]\right\}\;.
\end{eqnarray}
For the Maxwell field, the energy momentum tensor is given by the expression
\begin{eqnarray}
\mathcal{T}^{\,\mu}_{\beta}=\frac{1}{4\pi}\left[\frac{1}{4}\delta^{\mu}_{\beta}F^{\sigma\gamma}F_{\sigma\gamma}-F^{\mu\sigma}F_{\beta\sigma}\right]\label{tme}\; ,
\end{eqnarray}
where $F_{\mu\nu}=\partial_{\mu}A_{\nu}-\partial_{\nu}A_{\mu}$ is the Maxwell tensor and $A_{\mu}$ the electromagnetic quadri-potential. 

\section{\large  Charged black hole model}\label{sec3}
We assume that charged static spherically symmetric black hole is described by the following metric:
\hspace{0,2cm}
\begin{equation}
dS^2=e^{a(r)}dt^2-e^{b(r)}dr^2-r^2\left[d\theta^{2}+\sin^{2}\left(\theta\right)d\phi^{2}\right]\label{ltb}\;,
\end{equation}
where the metric parameters $\{a(r),b(r)\}$ are assumed to be functions of radial coordinate $r$ and are not time dependent. This is a consequence of Birkhoff theorem 
 which governs the generalized gravity with arbitrary choice of tetrads \cite{han}. In general, this statement shows how static distributions of matter, behaving as a static spacetime manifold (Riemannian or Weitzenbock), is a central key of any theory of gravity. In absence of this theorem, solar system tests fail and there is no direct way to measure the metric parameters  of that theory by using observable parameters. 
\par
From (\ref{ltb}) we have different choices of tetrads. The diagonal tetrads restrict the algebraic expression of $f(T)$ to the teleparallel linear form \cite{Boehmer}.   For constructing a good set of non-diagonal tetrads, it is just necessary to follow the idea developed in the references \cite{maluf2,maluf3}. The degree of freedom can be fixed by choosing the component $ e_{0}^{\;\;\mu} = u^{\mu}$, where $u^{\mu}$ is a quadri-velocity of an observer. The other components are chosen to be oriented along the directions of the Cartesian axes $\{x,y,z\}$. 
\par
Therefore, according to the technique previously proposed, we choose the following non-diagonal tetrad basis in which we perform a local Lorentz transformation on the diagonal basis, appropriately \cite{manuel3}:
\begin{eqnarray}\label{nontet}
\{e^{a}_{\;\;\mu}\}=\left[\begin{array}{cccc}
e^{a/2}&0&0&0\\
0&e^{b/2}\sin\theta\cos\phi & r\cos\theta\cos\phi &-r\sin\theta\sin\phi\\
0&e^{b/2}\sin\theta\sin\phi &
r\cos\theta\sin\phi &r\sin\theta
\cos\phi  \\
0&e^{b/2}\cos\theta &-r\sin\theta  &0
\end{array}\right]\;,
\end{eqnarray}  
where we define the determinant of the tetrad by  $e=det[e^{a}_{\;\;\mu}]=e^{(a+b)/2}r^2\sin\theta$. The non null components of the torsion tensor  (\ref{tor}) are
\begin{eqnarray}
T^{0}_{\;\;10}=\frac{a'}{2}\,,\;\;\;T^{2}_{\;\;21}=T^{3}_{\;\;31}=\frac{e^{b/2}-1}{r}\,\,,
\end{eqnarray}
while the non-null components of the contorsion tensor read
\begin{eqnarray}
K_{\;\;\;\;0}^{10}=\frac{a'e^{-b}}{2}\,,\;\;\;K_{\;\;\;\;1}^{22}=K_{\;\;\;\;1}^{33}=\frac{e^{-b}(e^{b/2}-1)}{r}\,\,.
\end{eqnarray}
The non-null components of the tensor $S_{\alpha}^{\;\;\mu\nu}$ can be computed, giving
\begin{eqnarray}
S_{0}^{\;\;01}=\frac{e^{-b}(e^{b/2}-1)}{r}\,,\;\;S_{2}^{\;\;12}=S_{3}^{\;\;13}=\frac{e^{-b}\left(a'r-2e^{b/2}+2\right)}{4r}\,.
\end{eqnarray}
\par
From the definition of the torsion scalar (\ref{t1}), one gets
\begin{equation}
T=\frac{2}{r}\left[-\left(a'+\frac{2}{r}\right)e^{-b/2}+\left(a'+\frac{1}{r}\right)e^{-b}+\frac{1}{r}\right]  \label{te}\,.
\end{equation}
Note that, here, with a general form of the metric, $T$ is not constant. 
\par
For the charged black hole configurations we need an additional Maxwell field (static). For the case of  static electric potential $A_{\mu}=[A_{0}(r),0,0,0]$, one has a unique component for Maxwell tensor, $F_{10}=\partial_r A_{0}(r)$. The Maxwell equations are
\begin{equation}
\nabla _{\mu}F^{\mu\nu}=0\label{me}\,,
\end{equation}
whose solution provides
\begin{eqnarray}
A_{0}(r)=\frac{q}{r}\;e^{(a+b)/2}\,,
\end{eqnarray}
where $q$ is the electric charge.
\par
From all the calculations above, we can establish the equations of motion from (\ref{em}), with (\ref{tme}), as
\begin{eqnarray}
&&2\frac{e^{-b}}{r}\left(e^{b/2}-1\right)T'f_{TT}+\frac{e^{-b}}{r^2}\left[b'r+\left(e^{b/2}-1\right)\left(a'r+2\right)\right]f_T+\frac{f}{2}=\frac{q^2}{r^4}\label{eq1}\,,\\
&&\frac{e^{-b}}{r^2}\left[\left(e^{b/2}-2\right)a'r
+2\left(e^{b/2}-1\right)\right]f_T+\frac{f}{2}=\frac{q^2}{r^4}\label{eq2}\,,\\
&&\frac{e^{-b}}{2r}\left[a'r+2\left(1-e^{b/2}\right)\right]T'f_{TT}+
\frac{e^{-b}}{4r^2}\Big[\left(a'b'-2a''-a'^2\right)r^2+\nonumber\\
&&+\left(2b'+4a'e^{b/2}-6a'\right)r-
4e^{b}+8e^{b/2}-4\Big]f_{T}+\frac{f}{2}=-\frac{q^2}{r^4}\label{eq3}\,.
\end{eqnarray}
\par
The system of field equations (\ref{eq1})-(\ref{eq2}) is a closed system for three unknown functions $\{a,b,f(T)\}$ highly coupled  in non linear forms, yielding a stiff system. For finding the possible exact solution of this system we need to have in hand all the three functions by solving simultaneously the system analytically. In fact, this is a very hard work and as a simple case we are only able to fix the form of $f(T)$ and solve the system analytically. Because we do not know how the boundary conditions change on the metric functions, the construction of numerical solutions is also difficult. In the next sections, first, we will  examine the system for $f(T)=T$, which corresponds to the TEGR, and later, we also will  show how the RN solution appears. 
\par 
Next, we find a family of exact solutions for a viable model of $f(T)$.

\subsection{Recovering the usual Reissner-Nordstrom-de Sitter (RN-dS) or RN-AdS cases}\label{subsec3.1}
In order to confirm the consistency of the theory, we propose to search whether the usual Reissner-Nordstrom-de Sitter (or RN-AdS) case may be recovered from the teleparallel theory, i.e. $f(T)=T-2\Lambda$. To do so, let us first perform the subtraction of (\ref{eq1}) from (\ref{eq2}),  getting after simplification the following equation
\begin{eqnarray}
\frac{e^{-b}}{r}\left( a'+b' \right)=0\,\,.\label{eq4}
\end{eqnarray}
Without loss for generality, one may set $a(r)=-b(r)$\footnote{This choice is commonly called quasi-global coordinate, for which, in GR, the equations of motion are independent on the gauge \cite{galtsov}.
}. Indeed, we can write $a(r)=-b(r)+a_0$ but it is straightforward to combine $a_0$ in a redefinition of time coordinate $t$. So, without loss of generality we put $a_0=0$. The physical meaning of this equality is related to the fact that we can interpret metric function $a$ as the potential of Newtonian form if we go to the non relativistic regime in which we approximate $g_{00}\sim1-2\Phi_G/c^2$,  $\Phi_G$ being Newtonian potential. If we change $a\rightarrow a+a_0$, nothing changes   because there is a gauge freedom for scalar Newton's potential.
\par
Therefore, Eq.~(\ref{eq3}) becomes
\begin{eqnarray}
e^{a}\left[\frac{a'^2+a''}{2}+\frac{a'}{r}\right]-\frac{q^2}{r^4}+\Lambda=0\,\,,\label{eq5}
\end{eqnarray}
which is the unique independent equation. By solving Eq.~(\ref{eq5}), one gets
\begin{eqnarray}
a(r)=\ln{\left(\frac{q^2}{r^2}+\frac{C_1}{r}+C_2-\frac{\Lambda}{3}r^2\right)}\,\,,
\end{eqnarray}
where $C_1$ and $C_2$ are integration constants. The constant $C_1$ is found by linearising the metric and comparing it at the Newtonian limit, yielding $C_1=-2M$, whereas the constant $C_2$ is obtained by assuming the Minkowskian limit (for $\Lambda=0$), obtaining $C_2=1$. Therefore, one gets
\begin{eqnarray}
a(r)=-b(r)&=&\ln{\left( 1-\frac{2M}{r}+\frac{q^2}{r^2}-\frac{\Lambda}{3}r^2\right)}\,,\label{RNS}
\end{eqnarray}
where $M$ is the mass of the black hole. 
\par
Note that Wang \cite{Wang:2011xf} has obtained the same solution, however with a choice different from the non-diagonal tetrads one. The strong difference between our analysis and that of Wang is that, the choice made by Wang leads to a restriction on the functional form of $f(T)$, in the case where the torsion scalar depends on the radial coordinate $r$, within a linear dependence on $T$. This is shown from $f_{TT}T^{\prime}=0$, which is Eq.~(52) of \cite{Wang:2011xf}, yielding black hole solutions. The problem with this analysis is that the constraint equation forces to two restricted possibilities: $a-$ $f_{TT}=0$, which leads to the linear case; $b-$ $T^{\prime}=0$, leading to the constant torsion. But in our approach to charged solutions, we no longer have restriction on the functional form of $f(T)$, as can be seen in (\ref{eq1})-(\ref{eq3}). With our choice of tetrads, we have various possibilities for the functional form of $f(T)$.
\par
For $\Lambda=0$, we get the teleparallel version of the Reissner-Nordstrom solution of the GR, because with
\cite{sotiriou}
\begin{eqnarray}
R=-T-2\nabla^{\mu}T^{\nu}_{\;\;\mu\nu}\label{scalarR}\,,
\end{eqnarray}
the curvature scalar for the solution given by (\ref{RNS}), reproduces very well the RN case with $R=0$ in (\ref{scalarR}). As the Maxwell energy-momentum tensor in (\ref{tme}) has a vanishing trace, in GR, we must get $R=0$ for RN solution. Here we have the same result, but the non-null function $T(r)$ is combined with the second term of the right hand side in (\ref{scalarR}),  cancelling identically. 
\par
We then re-obtain the RN-dS (RN-AdS) solutions, for $\Lambda=0$, the RN, for $q=0$, the Schwarzschild-de Sitter (S-dS) or S-AdS solution. The vacuum case is recovered in the limit of a vanishing charge and cosmological constant, $q\rightarrow 0$ and $\Lambda\rightarrow 0$.
\par
Here we obtain a new interpretation for vacuum solution in the teleparallel theory as the analogous solution to the Schawarzschild one in GR. The argument used by Ferraro and Fiorini in \cite{ferraro1}, according to what the torsion scalar should be zero or constant for a vacuum solution, following the analogy of GR, is still valid. But we also observe here that there is a new possibility of interpreting a vacuum solution, which is the mutual cancellation between the two terms of (\ref{scalarR}). This maintains the curvature scalar to zero and reflects the exact analogy of the teleparallel theory equivalent to GR. The teleparallel equivalent to GR is just dynamically equivalent to GR, and this lets us free to choose tetrads that may, or not, vanish the torsion scalar, since on the same way, the equations that govern the dynamics of the theory continue being equivalent. 
\par
As previously shown in \cite{manusug2}, $f(T)$ gravity has equations of motion depending on the choice of the set of tetrads. We also know that it is a theory for which the invariance under local Lorentz transformation no longer is realized \cite{manusug1}. Therefore,  as the relation between the torsion scalar and the curvature scalar has a term which is not invariant under the Lorentz transformations, it is natural to think that there is more than one possibility to re-obtain the known GR solutions by analogy.

\section{Reconstruction for Maxwell-f(T) type black holes}\label{sec4}
A recent and important mechanism for the modified theories of gravity is the reconstruction scheme  of the functional form of the related action. There are various examples of modified theories, namely, $f(R)$ \cite{fdeR}, $f(G)$ \cite{fdeG}, $f(R,\mathcal{T})$ \cite{fdeRT}, where
$R$, $G$ and $\mathcal{T}$ are the curvature scalar, the Gauss-Bonnet invariant and the trace of the energy momentum tensor, respectively, and also $f(T)$ under consideration in the present paper. In the case of static solutions, there are several solutions obtained by this method \cite{Wang:2011xf,manuel2}.
\par
As the equations of motion (\ref{eq1})-(\ref{eq3}) are highly nonlinear and coupled, the usual methods of resolution do not apply here. Due to the great difficulty in obtaining new solutions to the equations of motion, in this section, we will take into account a simplification for these equations, and therefore, being able to find suitable interpretation to $f(T)$ solutions of Maxwell type. A very typical consideration in obtaining solutions to the $f(T)$ theory is taking the matter content as being directly proportional to the algebraic function $f(T)$ and it derivative \cite{manuel2}. This feature can be directly seen from the equations (\ref{em}).

\par
Hence,  we take the first term in brackets on the left hand side of (\ref{eq2}), being identically null. This yields  
\begin{eqnarray}
&&f(T)=16\pi\mathcal{T}^{1}_{1}=q^2/r^4\label{f2}\,,\\
&&\left(e^{b/2}-2\right)a'r
+2\left(e^{b/2}-1\right)=0\label{sp1}\,.
\end{eqnarray}
We then determine $b(r)$ in terms of the derivative of $a(r)$, using (\ref{sp1}), and get  
\begin{equation}
b(r)=2\ln\left[\frac{2 (1 + ra')}{(2 + ra')}\right]\label{b1}\,.
\end{equation}      
One can then find  $(f_T,f_{TT})$, only in terms of  $r$ and $a'(r)$, in (\ref{eq1}) and (\ref{eq3}), such that they satisfy these equations. Doing this, we can infer a general $f(T)$ solutions of Maxwell type for $a(r)$, such that all equations are satisfied. One can make use of the following  ansatz
\begin{equation}
a(r)=\ln\left[1-\frac{2M}{r}+\frac{q^2}{r^2}-\frac{\Lambda}{3}r^2+a_1(r)\right]\label{a2}\,,
\end{equation}  
which, for $a_1(r)=0$, gives rise to 
\begin{eqnarray}
&&\exp[b(r)]=\frac{9(q^2-r^2+\Lambda r^4)^2}{r^2(3M-3r+2\Lambda r^3)^2}\label{b3} \,,\\
&&T(r)= -\frac{2(3q^2-3Mr+\Lambda r^4)^2}{3r^2(q^2-r^2+\Lambda r^4)[-3q^2+r(6M-3r+\Lambda r^3)]}\,.\label{te1}
\end{eqnarray}
We can also add the term $a_1(r)=a_0r$ which is interpreted as the Rindler acceleration for large scales \cite{grumiller}, leading to  
\begin{eqnarray}
&&\exp[b(r)]=\frac{36[q^2+r^2(-1-2a_0 r+\Lambda r^2)]^2}{r^2[6M+r(-6-9a_0 r+4\Lambda r^2)]^2}\label{b4} \,,\\
&&T(r)=-\frac{[6q^2-6Mr+r^3(-3a_0+2\Lambda r)]^2}{6r^2[q^2+r^2(-1-2a_0 r+\Lambda r^2)][-3q^2+r(6M-3r-3a_0 r^2+\Lambda r^3)]} \,.\label{te2}
\end{eqnarray}
Here, we see that by setting $a_0=0$ in (\ref{b4}) and (\ref{te2}), we regain (\ref{b3}) and (\ref{te1}), meaning that the last solution is a generalization of the first one. But they possess quite different properties, mainly, regarding the functional form of each $f(T)$ model. 
\par
We can even do a parametric plot, where the parameter is the radial coordinate $r$. From (\ref{f2}) we know that $f(r)=q^2/r^4$, and also we have the solution (\ref{te1}) for $T(r)$. Using $r$ as a parameter, we can represent the algebraic function $f(T)$ at Figure~$1$\footnote{ 
We remove the condition $T(r)=0$, where the graph are represented by dashed lines  in the two figures.}.  Taking the same algebraic function $f(T)$ with  our ansatz (\ref{f2}) and the solution (\ref{te2}), we describe parametrically $f(T)$ at Figure~$2$.

\begin{figure}[h]
\begin{center}
\includegraphics[height=5cm,width=0.4\textwidth, angle=0]{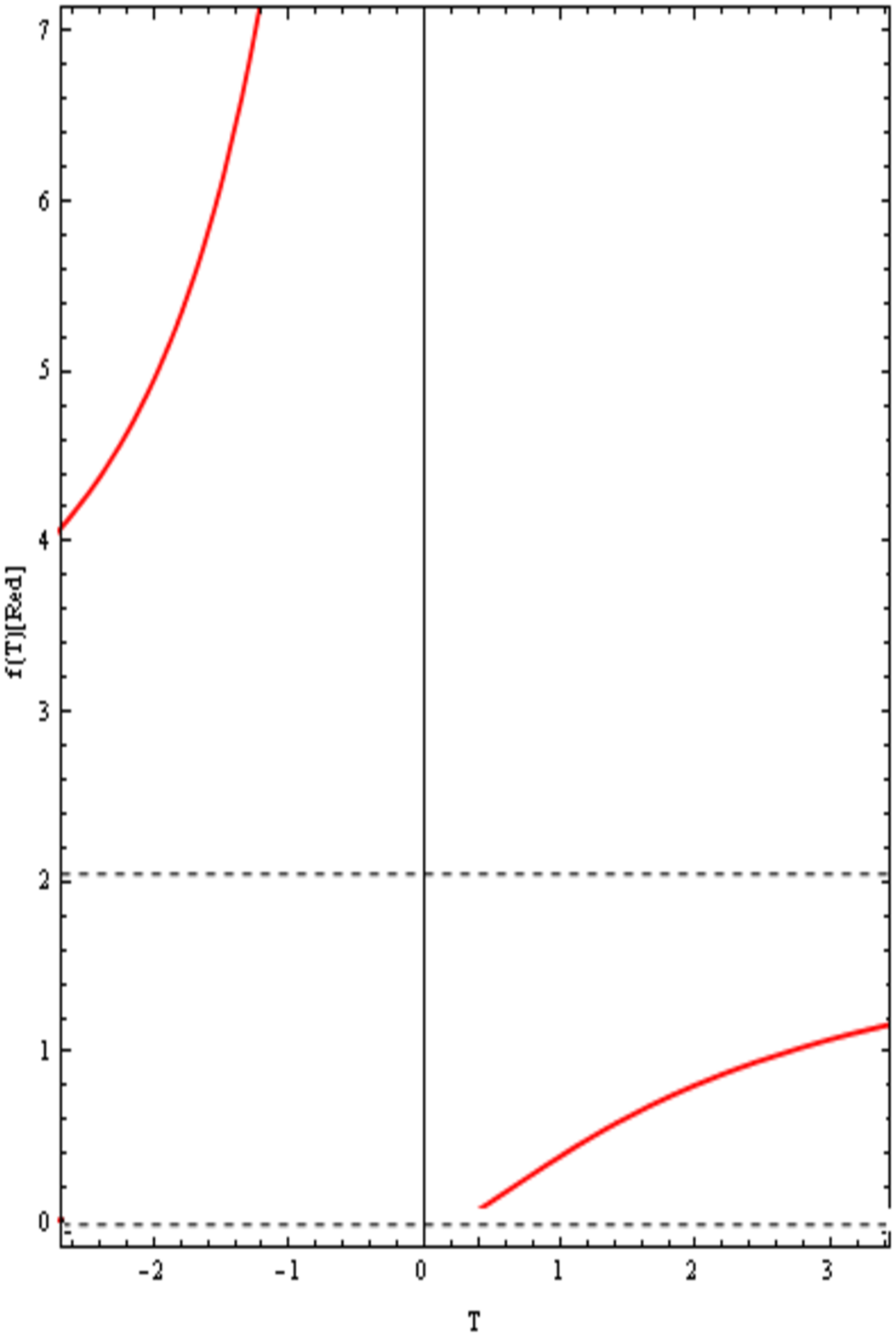}
\caption{{\small Parametric plot of $f(T)\times T$ for $a_1(r)=0,q=1,M=2,\Lambda=-0.01$ and $r\in[0.05,5]$.}}
\end{center}
\end{figure}
\begin{figure}[h]
\begin{center}
\includegraphics[height=5cm,width=0.4\textwidth, angle=0]{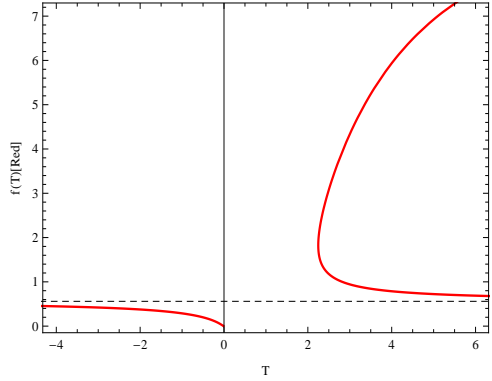}
\caption{{\small Parametric plot of $f(T)\times T$ for $a_1(r)=a_0 r,q=1,M=2,\Lambda=-0.01,a_0=1$ and $r\in[0.05,5]$.}}
\end{center}
\end{figure} 
\par
We can see from the two figures that we are dealing with the cases of nonlinear algebraic $f(T)$ function. We cannot reconstruct algebraically this function, since the equations (\ref{te1}) and (\ref{te2}) cannot be inverted to obtain $r(T)$ and substituting  in (\ref{f2}). Hence, our parametric analysis is necessary. 

\section{Why no black hole solution exists with $a'+b'=0$  in the non TEGR gravity and with $T\neq T_0$  ?}\label{sec5}
In the previous sections we derived some exact black hole solutions in the non constant torsion scalar case. Now, in this section, we prove that very useful and suitable gauge $a'+b'=0$ which simplifies calculations, is only accessible in the cases of $T=T_0,\ \ f_{T}(T_0)=0$, or only when we study TEGR action. This means that if we try to solve the system of equations (\ref{eq1}-\ref{eq3}) in another cases, not in the two mentioned cases (constant torsion of TEGR), we are not able to use the simplification solver ``key'' $a'+b'=0$. So, there is no Schwarzschild like solution for $f(T)$ with variable $T$ or far away from TEGR. This is a very significant result, because some works studied black holes with special case of gauge $a'+b'=0$ without notice this important fact. Explicitly,  we put out the following theorem:
\par
\emph{Theorem}: \textit{For a non TEGR case and without constant torsion scalar, i.e. with $T\neq T_0$, the metric functions $\{a,b\}$ cannot satisfy the reducible constraint $a'+b'=0$. So, it is impossible to solve black hole equations of motion in $f(T)$ with the gauge $a(r)=-b(r)$}.
\par
\emph{Proof.} By subtracting one of the equations (\ref{eq1}) and (\ref{eq2}) from other, in general, one gets
\begin{eqnarray}
\frac{2(f_T)'}{f_T}+\frac{a'+b'}{e^{b/2}-1}=0\label{proof1}\,.
\end{eqnarray}
This equation is valid without any additional assumption on the form of $f(T)$ or $T$. From (\ref{proof1}), we integrate and find explicitly
\begin{eqnarray}
f_T=\alpha\exp\{-\frac{1}{2}\int{\frac{a'+b'}{e^{b/2}-1}dr}\}\label{proof2}\,.
\end{eqnarray}
We analyse (\ref{proof2}) in the following cases.
\par 
TEGR case: with TEGR, we have: $f_T=1\Longrightarrow (f_T)'=0\Longleftrightarrow a+b=\text{constant}$. For this reason we can recover the RN spacetime in the TEGR limit  using the field equations, successfully.
\par
Constant torsion scalar case, $T=T_0$: in this case,  $T'=0\Longrightarrow (f_T)'=0\Longleftrightarrow f_{T}(T_0)(a'+b')=0$. Now, we have two possibilities:
\par 
A - First: if $f_{T}(T_0)=0$, then (\ref{proof1}) becomes an identity; so our preposition is valid;
\par 
B - Second: if $f_{T}(T_0)\neq0$, we consequently  have $a'+b'=0$, which again proves our theorem.
\par
So, in general, for a generic form of $f(T)$ and with a non constant torsion $T\neq T_0$, we loss simplification of gauge fixing $a'+b'=0$, i.e. Schwarzschild like metric.
\par
For this reason, it is a very hard task to find exact solutions for $T\neq T_0$. A simple reason to this is that, if we substitute (\ref{proof2}) in the new set of equations (\ref{eq1}+\ref{eq3}) and (\ref{eq2}+\ref{eq3}), we find the following system of coupled differential equations for $\{a,b,f(T)\}$:
\begin{eqnarray}
b' = h(a,b,a',a'',f;r) \,,\\
a'' =k(a,b,a',b',f;r)\,.
\end{eqnarray}
The system is highly non-linear and we cannot easily solve it analytically. Note that here 
$$f=f(T)=f(a',b',b;r)\,,$$
and satisfies (\ref{proof2}). So, we have a system of three unknown functions. It is possible to eliminate $f(T)$ from the above equations and obtain a pair of differential equations for $\{a,b\}$ and solve them analytically.
\par
For example, considering a very simple ansatz $a=kb,\ \ k\neq-1$, from (\ref{proof2}), we find
\begin{eqnarray}
&&f_{T}=\alpha (1-e^{-b/2})^{-(k+1)}\Longrightarrow f'=\frac{df}{dr}=\alpha T'(1-e^{-b/2})^{-(k+1)}\nonumber\\
&&\Longrightarrow f=\alpha\int^r{T'(x)(1-e^{-b(x)/2})^{-(k+1)}}dx \label{f}.
\end{eqnarray}
We substitute (\ref{f}) in (\ref{eq2}) for $k=1,\alpha=1$ and $q=1/3$, and find
\begin{eqnarray}
&&\left(-2+9 r^2\right) w^5+\left(6-9 r^2\right) w^4-3 \left(2+3 r^2\right) w^3+ \left(2+9 r^2+9 r^4 w''\right)w^2\nonumber\\
&&-9 r^4 \left(w'^2+w''\right)w-9 r^4 w'^2=0\,,
\end{eqnarray}
where we have $w(r)=\exp[b(r)/2]$. The equation can be solved numerically by appropriate boundary conditions. For example we plot the solution for a set of parameters and this shows that when $r\rightarrow+\infty$ we have $w\rightarrow+\infty$. Consequently, the spacetime is non-asymptotically flat (NAF) (see Fig. 3). Also, we have an event horizon in $r=r_H=1$ ($g_{00}(r_H)=w(r_H)=0$). Our numerical solution resembles a specific class of NAF black holes, for $\{ \gamma=1.28,r_0=b=r_H=1\}$ in ($2.8$) and ($2.9$) of \cite{gerard}. We plot $r^{\gamma}(r-1)$, in Fig. 1, in comparison to $w(r)$, where our solution is more convex downward. 
\begin{figure}[h]
\begin{center}
\includegraphics[width=0.4\textwidth, angle=0]{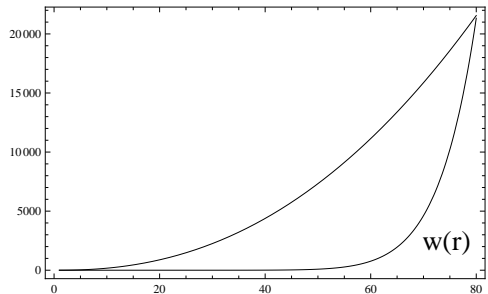}
\caption{{\small Plot of metric  function $w(r)$ and $r^{\gamma}(r-1)$ for $\alpha=1,k=1,q=1/3,w(1)=0.0001,w'(1)=2$ and $\gamma=1.28$ versus the radial coordinate $r$. The graphs started from horizon ($r_H=1$) to infinity and show that spacetime is non-asymptotically flat. Our solution is more convex downward.}}
\end{center}
\end{figure}
This is the second example of a solution of charged black hole with a non-linear $f(T)$  (see the first in \cite{Wang:2011xf}). Here, we cannot reconstruct the algebraic function $f(T)$. We will develop this interesting feature in a future work.

\section{About no-go theorem for charged black hole in f(T)}\label{sec6}
We consider a diagonal frame of tetrads in which we minimally coupled a Maxwell field tensor $F$ to gravity via torsion. It is easy to show that no-go theorem in diagonal frame is satisfied identically. In \cite{saridakis2}, a version of no-go theorem stated that there is no possibility to have two non vanishing components of electric (magnetic) field simultaneously. For example, only radial of azimuthal component of field (electric or magnetic) exists and satisfies all field equations appropriately, and if we insert both components, a serious inconsistency happens. In the previous section, we proved a simple but very useful theorem on the non existence of Schwarzschild gauge in a generic model of $f(T)$ with $T\neq\text{constant}$. For electric field, a no-go theorem has been proved and in three dimensional cases an explicit proof has been presented in \cite{saridakis1}. But the situation in non-diagonal frame like our case is so complicated and different. Our aim in this section is to check the validity of a type of no-go theorem for our model in the non-diagonal case. Here, the statement of no-go theorem is given as follows. 
\par
\textit{Theorem}:\emph{ There is no consistent metric in the form of $g_{\mu\nu}=diag(e^{a(r)},-e^{-a(r)},-r^2\Omega_2)$ for charged Maxwell field non minimally coupled to torsion via $f(T)$ gravity}. The proof was given in the previous section.
\par
So, we also presented a ``new'' no-go theorem for $f(T)$.

\section{Final Remarks}\label{sec6}
The charged black hole solutions in generalized teleparallel gravity models in Weitzenbock spacetime are revisited in a non-diagonal tetrads basis in $4D$. As advantage of this non-diagonal components, we avoid the restriction $f_{TT}=0$ which leads to TEGR where the result is well known as Reissner-Nordstrom-dS (RN-AdS) spacetime. 
\par
We derived field equations of a general $f(T)$ gravity in the first steps. Then, by assuming a non-diagonal tetrads basis in static coordinates we derived the full system of field equations. We proved an important theorem which shows why the common Schwarzschild form of metric in new tetrads formalism cannot be used for a general $f(T)$ gravity  without $f_{TT}=0$. According to this theorem we concluded that charged black holes in a general $f(T)$ is a complicated system of coupled differential equations. As examples, we separately analyse the cases of TEGR, recovering our field equations, re-obtaining the usual solutions of GR, as RN-dS (RN-AdS), RN, S-dS (S-AdS) and Schwarzschild. 
\par
The important point here is that, charged static black hole in TEGR has a non constant torsion scalar $T$. This holds in GR in spite of the spacetime having a vanishing curvature scalar, $R=0$. We show that  the analogy is still possible because the two terms that result from the curvature scalar can mutually cancel each other, then, leading to a new interpretation for these solutions in $f(T)$ gravity. We tested and proved a no-go theorem based on the inconsistency of charged black holes in non-diagonal case with a Schwarzschild type metric. Our work exposes new features of $f(T)$ gravity as an alternative to GR.

\vspace{0,5cm}
{\bf Acknowledgement:}  M. E. Rodrigues wishes to thank PPGF of the UFPA for the hospitality during the development of this work and and thanks CNPq for partial financial support.

\end{document}